\documentclass[12pt]{article}
\usepackage{psfig}

\input epsf
\newcount\figno
\def\fig#1#2#3{
\par\begingroup\parindent=0pt\leftskip=1cm\rightskip=1cm\parindent=0pt
\baselineskip=11pt
\global\advance\figno by 1
\epsfxsize=#3
\centerline{\epsfbox{#2}}
\vskip 12pt
{\bf Figure \the\figno:} #1\par
\endgroup\par
}
\def\figlabel#1{\xdef#1{\the\figno
\mbox{ }}}
\def\encadremath#1{\vbox{\hrule\hbox{\vrule\kern8pt\vbox{\kern8pt
\hbox{$\displaystyle #1$}\kern8pt}
\kern8pt\vrule}\hrule}}

\def\href#1#2{#2}
\def\beq{\begin{equation}}
\def\arccosh{\mbox{arccosh}}
\def\eeq{\end{equation}}

\def\drawbox#1#2{\hrule height#2pt
        \hbox{\vrule width#2pt height#1pt \kern#1pt
              \vrule width#2pt}
              \hrule height#2pt}

\def\Asym#1#2{\vcenter{\vbox{\drawbox{#1}{#2}
              \kern-#2pt       
              \drawbox{#1}{#2}}}}

\begin{document}
\begin{titlepage}

\begin{center}
\today
\hfill UW-PT-03/08\\
\hfill hep-th/0305192
\vspace{2cm}

{\Large\bf Auto-localization in de-Sitter space}

\vspace{2cm}
{\large
Andreas Karch }

\vspace{1cm}

{\it
{Department of Physics,
University of Washington,
Seattle, WA 98195, USA\\
[.4cm]}
}
({\tt karch@phys.washington.edu})\\

\end{center}

\vspace{1.5cm}

\begin{abstract}
We point out that gravity on dS$_n$ gives rise to a localized
graviton on
dS$_{n-1}$. This way one can derive a recursion relation
for the entropy of dS spaces, which might have interesting implications
for dS holography. In the same spirit we study
domain walls interpolating between dS spaces with different cosmological
constant. Our observation gives an easy way to calculate what fraction
of the total entropy can be accessed by an observer stuck on
the bubble wall.
\end{abstract}

\end{titlepage}

\section{Counting degrees of freedom in compactifications}

Consider a higher dimensional, or to be specific: a 
10-dimensional space-time of the product form dS$_5 \times M_5$ where $M_5$
is a compact space of volume $V$ and dS$_5$ is a de-Sitter
space of curvature radius $L$. At long distances gravity
can be described by an effective 5d theory, where the effective
5d Planck scale is given by
\beq
M_{pl,5d}^3 = V M_{pl,10d}^8.
\eeq
This spacetime has an entropy which is given by
\beq
\label{n}
S = \frac{1}{4} {\cal A}_3 L^3 V M_{pl,10d}^8
\eeq
where
\beq
{\cal A}_n=  2 \frac{ \pi^{\frac{n+1}{2}}}{\Gamma[\frac{n+1}{2}]}
\eeq
denotes the volume of the unit $n$-sphere.
The formula can be derived in two ways: in the full microscopic
geometry ${\cal A}_3 L^3 V$ is the horizon area.
In the effective long
distance description, ${\cal A}_3 L^3$ is the usual 5d dS horizon area in
the theory with effective Planck scale $V M_{pl,10d}^8$.
So both in 10d and in 5d the entropy is the horizon area in Planck units.

One way to understand why the effective 5d counting has to reproduce
the right number, is that according to rather general argument, see
e.g. \cite{banks} (or see \cite{bousso} for a review),
the area of the horizon in Planck units is really a bound on the total number
of degrees of freedom accessible to any quantum theory containing gravity
in a dS spacetime.
The 10d theory
on dS$_5 \times M_5$ is just a particular realization
of a quantum theory on dS$_5$ containing the massless graviton and
in addition a tower of massive KK spin 2 particles.
So the entropy in the full ``microscopic'' 10d
description has to agree with number derived
from the 5d horizon area in 5d Planck units.

Indeed we expect this counting to still be true even if the volume
$V$ of $M_5$ is of the same order as the curvature radius $L$ of
the dS space. In this case 5d gravity would not be a good
description to describe the force law an observer in the dS$_5$
space would measure for test masses that a separated by distance
smaller than the size of the horizon. However the usual arguments 
still suggest that the accessible number of degrees of
freedom in this theory containing gravity and a tower of light
spin-2 KK modes are still finite and given by the 5d horizon area.
We will show that in the same spirit
one can derive an amusing recursion relation between the 
entropy of
dS spaces of various dimensions.
The formalism developed
can also be used to gain some insight into 
the entropy of bubble spacetimes.

\section{Auto-localization in dS space}

In this section we are going to establish that gravity
in pure dS space can be viewed as lower dimensional dS gravity
coupled to a KK-mode matter system.
In order to see a localized graviton in a mechanism
similar to the one employed by RS \cite{RS}, we first need
to write dS as a warped product. It turns out that the only
allowed spatial slicing in terms of maximally symmetric
codimension one spacetimes is a dS slicing

\begin{figure}
 \centerline{\psfig{figure=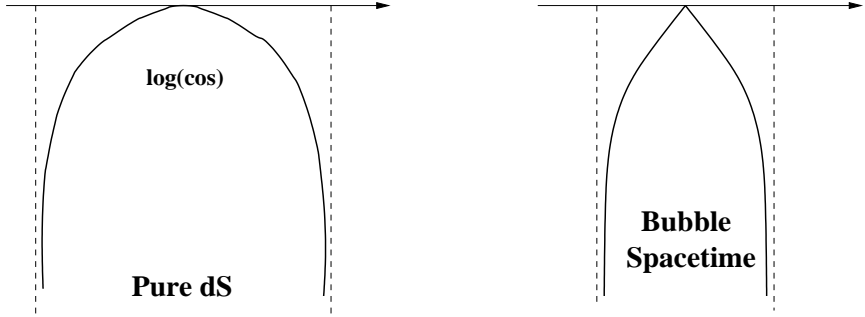,width=4.5in}}
 \caption{
Warpfactor for the de Sitter slicing of de Sitter space. Even
without any brane or defect we get a localized, normalizable
zero mode. Including a brane as usual pastes together
two regions of pure de Sitter.
}
\label{warps}
  \end{figure}

\beq
ds_{dS_d}^2 = e^{2 A(r)} ds^2_{dS_{d-1}} + dr^2
\eeq
where $dS_{d-1}$ has curvature radius $1/\sqrt{\Lambda}$.
The corresponding warpfactor is given by
\beq
A = \log(\sqrt{\Lambda} L \cos {\frac{r}{L} } ).
\eeq
Fixing $A(0)=0$ yields $\Lambda=\frac{1}{L^2}$, the $dS_{d-1}$ radius
at the turn around point is the same as the $dS_d$ radius.
Go to conformal form
\beq
ds_{dS_d}^2 = e^{2A(z)} \left ( ds^2_{dS_{d-1}} + dz^2 \right ).
\eeq
with
\beq z(r) = \frac{1}{\sqrt{\Lambda}} \arccosh \left (
\frac{1}{\cos(\frac{r}{L})} \right ). \eeq 
\beq e^{A(z)} = \frac{1
 }{\cosh ( z \sqrt{\Lambda} )} \eeq 
Using the formalism and conventions of \cite{dfgk} we can
translate the differential equations of linearized gravity
into an analog quantum mechanics with volcano potential
\beq V(z) =
\frac{(d-2)^2}{4} A''(z) + \frac{d-2}{2} A'(z)^2 = \Lambda \left (
\frac{(d-2)^2}{4} - \frac{d-2}{4} d \frac{1}{\cosh^2(z
\sqrt{\Lambda})}  \right ). \eeq

\begin{figure}
 \centerline{\psfig{figure=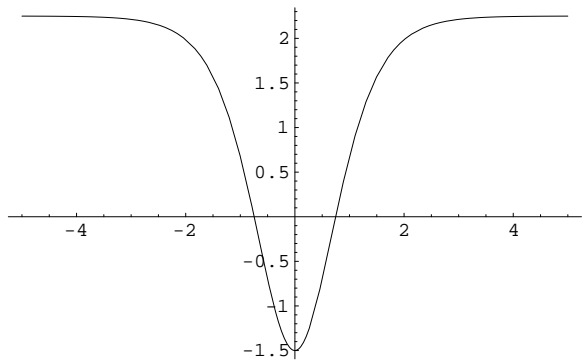,width=3.0in}}
 \caption{
Volcano potential for the dS slicing of pure dS.
This (and the other potentials shown) is plotted for $d=5$ and
$L=1$.
}
\label{ds}
  \end{figure}

This potential has a single bound state at zero energy, separated
from a continuum of modes separated by a mass gap of order $\frac{1}{L^2}$.
The zero mode solution 
\beq \psi=e^{\frac{d-2}{2} A(z)} =  \left  (
\cosh ( z \sqrt{\Lambda} ) \right )^{\frac{2-d}{2}} \eeq
is a normalizable solution to the wave equation with $E=0$. 
So the full theory can be rewritten in terms of a $d-1$ dimensional
graviton on the slice at $r=0$.

To relate the $d$-dimensional Planck scale to
the $d-1$ dimensional Planck scale we can use the fact
that the norm of the zero mode is effectively the volume
of the internal space.
Evaluating the norm
\beq V \equiv  \int dz |\psi(z)|^2 = \sqrt{\pi} 
\frac{\Gamma(\frac{d}{2}-1)}{\Gamma(\frac{d-1}{2})} L \eeq
we get
\beq M_{Pl,d-1}^{d-3} = V \; M_{Pl,d}^{d-2} = \frac{A_{d-2}}
{A_{d-3} } \;  L \;  M_{Pl,d}^{d-2} \eeq
and one can check that indeed the entropy of the $dS_{d-1}$ using
the Planck scale of the localized graviton is identical
to the entropy of the $dS_d$ using its original Planck scale:
\beq 4 S = A_{d-2} \;  M_{Pl,d}^{d-2} \; L^{d-2} =
         A_{d-3} \; M_{Pl,d-1}^{d-3} \; L^{d-3}. \eeq
Obviously the theory is not really lower dimensional. Even though
the continuum modes are separated by a mass gap, they contribute
at length scales smaller than the horizon radius, since their mass
is set by the dS radius. But still, as in the case of a
compactification, arguing that any theory of gravity
in de Sitter space, even
a theory containing a continuum of light spin-2 modes,
should not exceed the entropy of the horizon area in Planck units,
we can reproduce the exact entropy in terms of the
localized graviton\footnote{
One way to see why the integral had to work is that the $dS$ slicing of $dS$
gets inherited after Wick rotation from the usual way to write
the $n$-sphere in terms of $n-1$ spheres:
\beq
ds^2_{S^n} = \cos^2(\theta) ds^2_{S^{n-1}} + d \theta^2. \eeq
Evaluating the volume of this space leads to precisely the same
integral over theta we used to evaluate the norm above.}.

{\bf A Note on holography:}

So far we analyzed the purely classical theory of gravity
fluctuations in dS$_d$ space. We found that they can be described
in terms of a localized graviton on the central dS$_{d-1}$ slice
coupled to a KK-continuum that is separated from the
zero mode by a mass-gap of order the inverse dS curvature radius.
In the similar situation in AdS it has been established that the KK
continuum can be holographically 
replaced by a $d-1$ dimensional quantum field theory, still
coupled to the classical localized graviton \cite{lisaRS,gubRS,witRS,porrati}.
If this would be possible here as well, repeated application
would provide us with a
way to replace gravity in any dS by gravity in a 2d dS space (which
recently has been shown to be a tractable theory even on the quantum
level \cite{martinec1,martinec2}) coupled to a complicated quantum
field theory representing all the KK modes. Eventually
this 2d system could even be replaced by a matrix model. This might be
the best way to get a handle on holography in dS spaces.

A different proposal has been advanced in the past, known as the dS/CFT
correspondence \cite{andy,witDS} (see also 
\cite{pum,pum2}), where the holographic dual of dS
gravity is a euclidean
CFT living at the boundary at future or past infinity. The
validity of this picture has been questioned by many recent papers, in
particular \cite{willy1,willy2,susskind}. In this language our decomposition
would correspond to the euclidean version of the AdS/dCFT story
of \cite{kr,kr2,dfo,bachaso}. In this language 
what we described above is the analog of the ``halfway dual''
of \cite{porrati,br}, where only the KK-modes get replaced by a quantum
CFT coupled to the classical graviton localized on the brane.
In dS latter might prove to be the more powerful statement.

\section{dS domain walls}

\begin{figure}
 \centerline{\psfig{figure=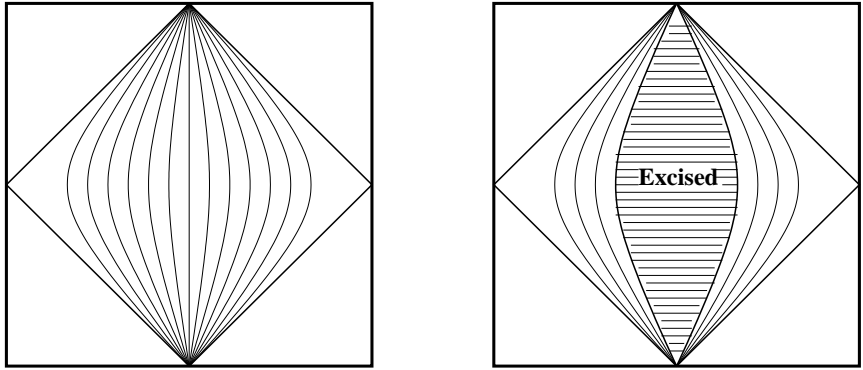,width=4.5in}}
 \caption{
Embedding of the dS slicing in dS. The brane cuts out
the shaded region in the diagram on the right.}
\label{dsslice}
  \end{figure}

It is easy to see what happens when we include domain walls
with a non-zero positive tension $\lambda$. 
They live at an $r=const.$ slice
such that
\beq \lambda  \propto A'|_{Left} - A'|_{Right} \eeq
These domain walls have received some recent attention \cite{eva}.
If one allows the cosmological constant to jump across
the brane (give them charge under a background $d$-form fieldstrength), 
they correspond to the bubbles of false vacuum
decay. In this section we will focus on the simplest
case of a neutral brane. For earlier studies of localization
of gravity on dS branes in dS see \cite{ortin,ito,o1,o2,o3}.

A zero tension ``phantom  brane'' lives at $r=0$, that is,
as we have seen above, without any brane pure de-Sitter localizes
a graviton on a de-Sitter slice in the ``center'' of dS. At finite tension
the brane lives at a finite $r_0$ (or $z_0$ in the conformal
coordinates) and the region with $|r|<r_0$
gets excised from space-time. The effective cosmological
constant on the brane is given by:
\beq
\label{cc}
\frac{1}{L_{d-1}^2} = \frac{1}{L^2} + \frac{1}{l^2} \eeq
where $l$ is the curvature radius associated with the cosmological
constant due to the brane tension $\lambda$.

\begin{figure}
 \centerline{\psfig{figure=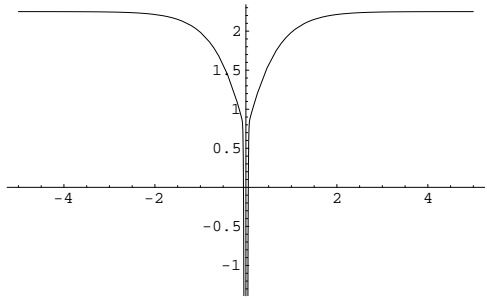,width=2.5in}}
 \caption{
Volcano potential for $z_0=1$ at $L=1$, $d=5$. 
}
\label{dsb}
  \end{figure}
\begin{figure}
 \centerline{\psfig{figure=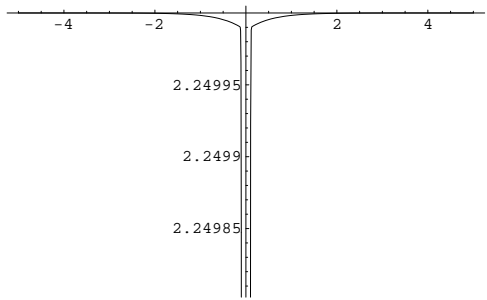,width=2.5in}}
 \caption{
Volcano potential for $z_0=7$ at $L=1$, $d=5$. 
}
\label{dsb2}
  \end{figure}

Volcano potentials for branes with increasing tension
are depicted in Figures \ref{dsb} and \ref{dsb2}. The structure
is always the same: a single boundstate zero mode separated
from a continuum by a mass gap 
\beq M^2_{gap} = \frac{(d-2)^2}{4 L_{d-1}^2}. \eeq
Note that it is the curvature radius of the brane dS
space that enters. In the extreme case of a $dS_{d-1}$ brane
in Minkowski space we just have the usual delta function
in an otherwise flat potential first studied in \cite{kaloper}.

It is interesting to ask, what happens to the entropy. As the
brane moves out to finite values of $r$, the curvature radius
of the lower-dimensional dS decreases with the warpfactor. But
at the same time the volume of space shrinks. The entropy
of the localized graviton on $dS_{d-1}$ is:
\beq 4S = A_{d-2} M_{Pl,d}^{d-3} V L_{d-1}^{d-3}   \eeq
where (setting $|\Lambda=\frac{1}{L^2}|$) 
\beq L_{d-1} = e^{A(z_0)} L < L \eeq
and 
\beq V=\int_{z_0}^{\infty} |\psi(z)|^2 < V_0
 =\int_{0}^{\infty} |\psi(z)|^2 . \eeq
That is, the entropy of the localized graviton is
bounded to be strictly less than the entropy of the higher dimensional de
Sitter space! 
Naively one would have expected that the entropy
of the excised region gets stored in a holographic fashion
as matter entropy on the brane. But lower dimensional gravity
predicts that the brane entropy is less than the
entropy of the original de Sitter space.
How shall we interpret this result? The entropy we calculated
is the entropy accessible to an observer on the brane. As usual in dS space,
parts of the brane will be hidden behind horizons. This entropy has
to be distinguished from the full "open string" entropy carried
by the brane, which featured prominently in \cite{eva}. There
are bulk observers who can access the full entropy of degrees of freedom
on the brane \cite{eva}. We have nothing new to say about the latter.

In the extreme case that the brane tension becomes much larger than the
bulk cc we recover the case of a dS brane in flat space. In this case
the bulk entropy goes to infinity when we keep the brane entropy finite.

\section{Newton Law on the brane?}

All we needed for our discussion of the previous sections
was to have a localized, normalizable zero mode in order to apply
the lower dimensional holographic bound. One may wonder whether in any
of the cases discussed above, gravity looks genuinely lower dimensional.
This is never the case. Note that even though all the KK-modes are massive,
the gap is always of order $1/L_{d-1}$. So the KK-modes can only
be neglected at distance scales larger than the horizon radius of the
lower dimensional dS space. At length scales accessible to
any observer on the brane, we always see the effect of the light, massive
KK continuum modes. Never do we get a lower dimensional Newton law.
But let's emphasize again, that this does not invalidate the entropy
counting in terms of the lower dimensional dS space.

In the same spirit as above we can similarly study domain walls
across which the bulk cc jumps, that is they are charged under
a $d$-form field strength. We are not restricted to having jumps
from dS into dS only, we can as well have jumps from dS into Minkowski or
AdS. In this case the warpfactor is $log(r)$ or $log(sinh(r))$
respectively instead
of the $log(cos(r))$. One can always find solutions that paste those
together like in the right diagram of Figure \ref{warps}, that is
in the ``up-down'' fashion. 

\begin{figure}
 \centerline{\psfig{figure=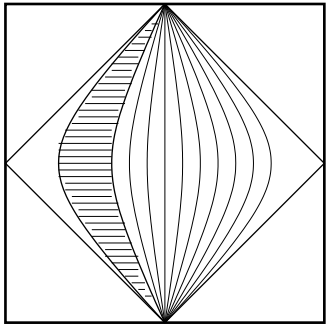,width=1.7in}}
 \caption{
Almost tensionless bubble 
interpolating between two dS spaces with different cosmological
constant.
}
\label{bubble}
  \end{figure}

The situation becomes more interesting when one considers 
``up-up'' spacetimes. Those are only possible for dS/dS bubbles.
This happens when the brane has charge but almost no tension.
The warpfactor than looks very close to the left diagram of 
Figure \ref{warps}
with a brane of center inducing a little additional jump. In this scenario
the localized graviton is not localized on the brane but it is localized
in the center of dS as in the no-brane case. The dS radius of the
localizing slice is hence always $L$, even as $L_{d-1}$ can become very small.

\section*{Acknowledgements} I would like to thank Dan Freedman
and Igor Klebanov for the organization of the string theory
workshop in Benasque in the summer of 2001 where most of this
work was done and I am grateful to all the participants for
useful discussions, in particular Ofer Aharony, Raphael Bousso and
Andy Strominger. Special thanks to Eva Silverstein for
getting me interested in this subject again and for stimulating
discussions about \cite{eva}. Thanks also to Josh Erlich and Ami Katz
for comments on the draft. This work was partially supported by the DOE
under contract DE-FGO3-96-ER40956.

\bibliography{ds}
\bibliographystyle{utphys}

\end{document}